\documentclass[aps,prl,twocolumn,superscriptaddress,floatfix,nofootinbib,showpacs,longbibliography,groupedaddress]{revtex4-1}

\usepackage{xcolor}
\usepackage[utf8]{inputenc}  
\usepackage[T1]{fontenc}     
\usepackage[british]{babel}  
\usepackage[sc,osf]{mathpazo}
\usepackage[scaled=0.86]{berasans}  
\usepackage[colorlinks=true, citecolor=blue, urlcolor=blue]{hyperref}  
\usepackage{graphicx} 
\usepackage[babel]{microtype}  
\usepackage{amsmath,amssymb,amsthm,bm,amsfonts,mathrsfs,bbm} 

\usepackage{xspace}  
\usepackage{xcolor}
\usepackage{multirow}
\usepackage{array}
\usepackage{bigstrut}
\usepackage{braket}
\usepackage{color}
\usepackage{natbib}
\usepackage{multirow}
\usepackage{mathtools}
\usepackage{float}
\usepackage[caption = false]{subfig}
\usepackage{xcolor,colortbl}
\usepackage{color}

\newcommand{\be}{\begin{equation}}
\newcommand{\ee}{\end{equation}}
\newcommand{\ba}{\begin{eqnarray}}
\newcommand{\ea}{\end{eqnarray}}

\def\>{\rangle}
\def\<{\langle}

\begin{document}
\title{On the Interpretation of Quantum Indistinguishability  : a No-Go Theorem}

\author{Anandamay Das Bhowmik}
\affiliation{Physics and Applied Mathematics Unit, Indian Statistical Institute, 203 BT Road, Kolkata, India.}

\author{Preeti Parashar}
\affiliation{Physics and Applied Mathematics Unit, Indian Statistical Institute, 203 BT Road, Kolkata, India.}


\begin{abstract}
Despite being the most fundamental object in quantum theory, physicists are yet to reach a consensus on the interpretation of a quantum wavefunction. In the broad class of realist approaches, quantum states are viewed as Liouville-like probability distributions over some space of physical variables where indistinguishabity of non-orthogonal states is attributed to overlaps between these distributions. Here we argue that such an interpretation of quantum indistinguishability is wrong. In particular, we show that quantum mechanical prediction of maximal violation of Mermin inequality in certain thought experiment is incompatible with all ontological interpretations for quantum theory where indistinguishability of non-orthonal quantum states is explained, even partially, in terms of overlap of their Liouville distributions.

\end{abstract}

\maketitle
{\it Introduction:--} The question of how to interpret a  quantum wavefunction still remains at the heart of many foundational debates in quantum theory. Einstein strongly advocated the view that quantum states should be thought of as probabilistic mixture of 'dispersion-free' real states so that quantum indeterminism can simply be explained as lack of knowledge over underlying deterministic states {\cite{EPR35}}. But in $1964$ John S. Bell gave a no-go theorem that shows such an interpretation is inconsistent with local causality \cite{Bell64}. Since then, no-go theorems have become a crucial approach to our understanding of quantum phenomena.  

Nearly a decade ago, a new set of no-go results \cite{Pusey12,Leifer14} has been derived from which arises a new foundational question on how to explain the indisntinguishablity of quantum states. Non-orthogonal quantum states cannot be distinguished with certainty in a single shot. Though this is often regarded as a distinctly quantum phenomenon, same is true also for classical probability distributions. If quantum states are viewed as Liouville-like classical probability distributions over states of reality (also called hidden variables or ontic states), two such quantum states whose distributions have nonzero overlap cannot be distinguished in a single shot experiment. In other words, quantum indistinguishability of any pair of non-orthogonal states can simply be explained in terms of the classical overlap of their respective distributions. We prove in this paper that such an interpretation of quantum indistinguishability, however reasonable it may sound, is incorrect. 

The first notable attack on the overlap-interpretation of quantum indistinguishability came from the seminal work by Pusey, Barrett and Rudolph (PBR) who have shown that distributions corresponding to non-orthogonal quantum states have zero overlap \cite{Pusey12}. However, their conclusion relies on an assumption that independently prepared systems have independent physical states (named as preparation independence assumption). Subsequently, several criticisms were raised regarding this assumption. \cite{Schlosshauer12,Emerson13,Ballentine14,Schlosshauer14} (also \cite{Leifer14}). Moreover, recent research reveals that in network scenario systems originated from independent sources yield correlations that cannot be reproduced by independent variables \cite{Branciard10,Renou19} - placing PBR's assumption into more questionable region. However,
the no-go theorem derived in this work requires no such controversial assumption in order to rule out theories that explain quantum indistinguishability by overlap between classical distributions. Thus the entire controversy regarding the validity of their assumption loses much relevance in this context. Though several other results were reported subsequently, all these results only put some bound on the amount of overlap \cite{Maroney12,Barrett14,Branciard14} and therefore, do not completely rule out the overlap-interpretation for quantum indistinguishability and also all these results apply to systems of dimension strictly greater than two \cite{Leifer13,Leifer14(1)}. In this paper, we test the validity of such an interpretation by considering a thought experiment and find that any model, in order to reproduce quantum predictions, must allow zero overlap between distributions for certain pairs of indistinguishable quantum states, thus making it necessary to have some additional postulate in the model in order to explain quantum indistinguishability.

{\it Framework:--} We consider ontological models \cite{Harrigan09,Spekkens05,Harrigan07}, where for every physical system there exists an ontological state space $\Lambda$ - the elements $\lambda$ of which are termed as ontic states. These ontic states should be thought of as the underlying
states of reality that a system can be in at a given time. The preparation of a quantum state $\ket{\psi}$ corresponds to a Liouville-like classical probability distribution $\mu(\lambda|\psi)$ over $\Lambda$, where each realization of the preparation $\ket{\psi}$ results in an ontic state $\lambda\in\Lambda$ sampled with probability measure $\mu(\lambda|\psi)$. The probability distribution $\mu(\lambda|\psi)$ is called the epistemic state (we shall often refer to this as Liouville distribution) associated with $\ket{\psi}$ and $\Lambda_{\ket{\psi}}:=\{\lambda\in\Lambda~|~\mu(\lambda|\psi)>0\}$ is called the ontic support of $\ket{\psi}$. For all $\ket{\psi}$, we must have $\int_{\Lambda_{\ket{\psi}}}d\lambda\mu(\lambda|\psi)=1$. 
If measurement of an observable $N$ is performed on a system, the possible outcomes belong to the set of eigenvalues $\tau_k$ with the associated eigenvectors $\ket{\tau_k}$, {\it i.e.}, $N=\sum_k \tau_k\ket{\tau_k}\bra{\tau_k}$. For notational simplicity, we are only describing the case for rank-1 projective measurements, although quantum theory allows more general measurement described by positive operator valued measure (POVM). If a system is in the ontic state $\lambda$ and observable $N$ is measured on it, the probability of obtaining the $i^{th}$ outcome is given by a response function $\xi(\tau_i|\lambda, N)\in[0,1]$. A generic ontological model keeps open the possibility for this outcome response to be contextual \cite{Spekkens05}; whenever measurement context is not important we will denote the response function simply as $\xi(\tau_i|\lambda)$. Define $\mbox{Core}[\xi(\psi|\lambda)]:=\{\lambda\in\Lambda~|~\xi(\psi|\lambda)=1\}$, then it immediately follows, $\Lambda_{\ket{\psi}}\subseteq\mbox{Core}[\xi(\psi|\lambda)]$. A {\it realistic} or {\it deterministic} ontological model demands $\xi(\tau_i|\lambda, N)\in\{0,1\}~\forall~i,\lambda,N$. The validity of an ontological model finally rests on whether it can successfully reproduce the Born rule at the operational level {\it i.e.} the model has to satisfy  $\int_{\Lambda}d\lambda\xi(\tau_i|\lambda,N)\mu(\lambda|\psi)=|\langle\tau_i|\psi\rangle|^2:=\mbox{Pr}(\tau_i|\psi)$ for all choice of measurement $N$ and preparation $\ket{\psi}$.

An ontological
model is $\psi$-epistemic if at least two distinct quantum states
$\ket{\psi}$ and $\ket{\phi}$ are described by distributions $\mu(\lambda|\psi)$ and $\mu(\lambda|\phi)$
such that their supports have an overlap of nonzero measure. The model is $\psi$-ontic otherwise. A model is maximally $\psi$-epistemic if the quantum overlap $|\langle\psi|\phi\rangle|^2$ between any two state $\ket{\psi}$ and $\ket{\phi}$ is explained completely in terms of the overlap between the corresponding distributions $\mu(\lambda|\psi)$ and $\mu(\lambda|\phi)$, {\it i.e.}, $\int_{\Lambda_{\ket{\phi}}}\mu(\lambda|\psi)d\lambda=|\langle\psi|\phi\rangle|^2$ \cite{Maroney12}. However, in general, $\int_{\Lambda_{\ket{\phi}}}d\lambda\mu(\lambda|\psi)=\int_{\Lambda_{\ket{\phi}}}d\lambda\xi(\phi|\lambda)\mu(\lambda|\psi)\leq \int_{\Lambda}d\lambda\xi(\phi|\lambda)\mu(\lambda|\psi)=|\langle\phi|\psi\rangle|^2$. The first equality is due to the fact that $\Lambda_{\ket{\phi}}\subseteq\mbox{Core}[\xi(\phi|\lambda)]$. The above inequality can be expressed as an equality as the following
\begin{equation}\label{deg}
\int_{\Lambda_{\ket{\phi}}}d\lambda\mu(\lambda|\psi)=\Omega(\phi,\psi)~|\langle\phi|\psi\rangle|^2
\end{equation}
where $\Omega(\phi,\psi)\in[0,1]$ measures the amount of overlap between $\mu(\lambda|\psi)$ and $\mu(\lambda|\phi)$. For a maximally $\psi$-epsitemic model, $\Omega(\phi,\psi)=1$ for all pairs of state. If $\Omega(\phi,\psi)=0$, there is no overlap between $\mu(\lambda|\psi)$ and $\mu(\lambda|\phi)$. If $\Omega(\phi,\psi)=0$ but $|\langle\phi|\psi\rangle|^2 \neq 0$, indistinguishability of $\ket{\psi}$ and $\ket{\phi}$ can no longer be explained in terms of overlap in their epistemic distributions.

    {\it Composite ontic state space :--} While each individual system has its own local ontic state space $\Lambda$, a composite system made up of a number of subsystems has a joint ontic state space. For instance, consider System-$A$, System-$B$ and System-$C$ , each having its local ontic state space $\Lambda^A$, $\Lambda^B$ and $\Lambda^C$ respectively. At this point, classical intuition may suggest that the joint ontic state space $\Lambda^{ABC}$ for the composite system-$ABC$ will simply be the Cartesian product of their individual local ontic state spaces {\it i.e.} $\Lambda^{ABC} = \Lambda^A \times \Lambda^B \times \Lambda^C$. However, if this is true, all joint ontic state $\lambda^{ABC} \in \Lambda^{ABC}$ can be decomposed as $\lambda^{ABC} \equiv (\lambda^A,\lambda^B,\lambda^C)$ where $\lambda^A \in \Lambda^A, \lambda^B \in \Lambda^B, \lambda^C \in \Lambda^C$ and any quantum preparation $\ket{\psi_{ABC}}$, being some distribution $\mu(\lambda^{ABC}|\psi_{ABC}) \equiv \mu(\lambda^A,\lambda^B,\lambda^C|\psi_{ABC})$ over $\Lambda^A \times \Lambda^B \times \Lambda^C$, should yield only local correlations that does not violate any kind of local realistic inequality. But the lesson from Bell's theorem is that there exist entangled states $\ket{\psi_{ABC}}$ that yield no-signalling correlations which cannot be reproduced by local hidden variable model \cite{Brunner14}. Therefore, $\Lambda^{ABC}$ cannot be only $\Lambda^A \times \Lambda^B \times \Lambda^C$. Violation of any local realistic inequality by some quantum state $\ket{\psi_{ABC}}$ necessitates some `nonlocal' ontic state space $\Lambda^{NL}$, in addition to $\Lambda^A \times \Lambda^B \times \Lambda^C$. Therefore, $\Lambda^{ABC} = (\Lambda^A \times \Lambda^B \times \Lambda^C) \cup \Lambda^{NL}$. Elements $\lambda^{ABC} \in \Lambda^{NL}$ are nonlocal, cannot be decomposed as  $(\lambda^A,\lambda^B,\lambda^C)$. Their presence in the ontic support $\Lambda_{\ket{\psi_{ABC}}}$ accounts for the observed nonlocality in the correlations $\ket{\psi_{ABC}}$ yields. For any entangled state $\ket{\psi_{ABC}}$ exhibiting nonlocal correlations, $\Lambda_{\ket{\psi_{ABC}}} \subseteq (\Lambda^A \times \Lambda^B \times \Lambda^C) \cup \Lambda^{NL}$ with $\Lambda_{\ket{\psi_{ABC}}} \cap \Lambda^{NL} \neq \emptyset$. On the other hand, if $\ket{\psi_{ABC}}$ is a product state, say $\ket{\psi_{ABC}} = \ket{\phi_A}\otimes \ket{\phi_B} \otimes \ket{\phi_C}$, for any $\lambda^{ABC}$ belonging in the ontic support $ \Lambda_{\ket{\phi_A} \ket{\phi_B} \ket{\phi_C}}$ , $\lambda^{ABC} = (\lambda^A,\lambda^B,\lambda^C)$ where $\lambda^A \in \Lambda_{\ket{\phi_A}} \subset \Lambda^A$, $\lambda^B \in \Lambda_{\ket{\phi_B}} \subset \Lambda^B$ and $\lambda^C \in \Lambda_{\ket{\phi_C}} \subset \Lambda^C$. Therefore, $ \Lambda_{\ket{\phi_A} \ket{\phi_B} \ket{\phi_C}} \subset (\Lambda^A \times \Lambda^B \times \Lambda^C)$ {\it i.e.} $\Lambda_{\ket{\phi_A} \ket{\phi_B} \ket{\phi_C}} \cap \Lambda^{NL} = \emptyset $. In other words, all $\lambda^{ABC}$ in the ontic support of product state $\ket{\phi_A}\otimes \ket{\phi_B} \otimes \ket{\phi_C}$ are local. Therefore we can write, $\mu(\lambda^{ABC}|\phi_A\otimes \phi_B \otimes \phi_C) \equiv \mu(\lambda^A,\lambda^B,\lambda^C|\phi_A\otimes \phi_B \otimes \phi_C)$. It is worth mentioning that PBR in their proof have made an assumption here : $\mu(\lambda^A,\lambda^B,\lambda^C|\phi_A\otimes \phi_B \otimes \phi_C) = \mu(\lambda^A|\phi^A)\mu(\lambda^B|\phi^B)\mu(\lambda^C|\phi^C)$\cite{Pusey12}. However, we do not need to assume this.

{\it Result:--} We now prove the no-go theorem. To this aim we consider a thought experiment with a quantum machine $\mathbb{M}$. The action of $\mathbb{M}$ is a unitary evolution $U_{\mathbb{M}}$ satisfying $U_{\mathbb{M}}\ket{0}_A\ket{r}_B\ket{r}_C = \ket{0}_A\ket{0}_B \ket{0}_C$ and $U_{\mathbb{M}}\ket{1}_A\ket{r}_B\ket{r}_C = \ket{1}_A\ket{1}_B \ket{1}_C$ where $\ket{r}$ is some fixed reference state. $\mathbb{M}$ has three input ports, $2$nd and $3$rd ports are fed with systems $B$ and $C$ respectively, prepared in the fixed reference state $\ket{r}$ and the $1$st one is fed with system $A$ prepared in state $\ket{0}$ or $\ket{1}$.

\begin{figure}[t!]
\centering
\includegraphics[width=0.35\textwidth]{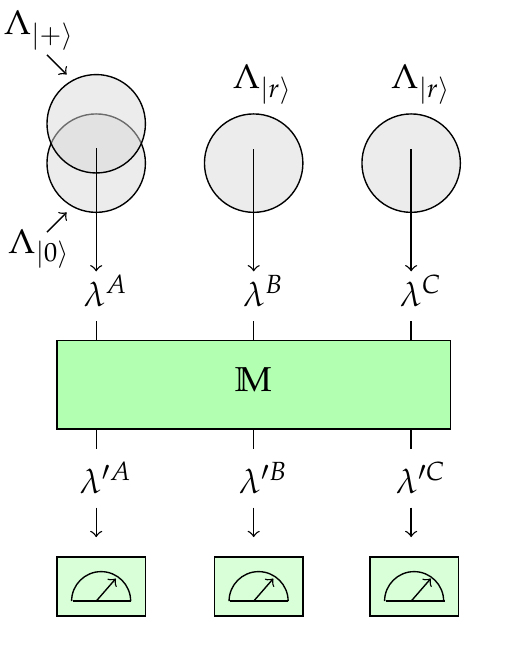}
\caption{(Color online) Schematic diagram of the thought experiment. Whenever $\lambda^A$ from $\Lambda_{\ket{0}}$ goes as input, machine acts on the input  $\lambda^{ABC} = (\lambda^A,\lambda^B,\lambda^C)$ to produce an output $\mathbb{M}[\lambda^A,\lambda^B,\lambda^C] = (\lambda'^A,\lambda'^B,\lambda'^C)$ where $\lambda'^A \in \Lambda_{\ket{0}} \subset \Lambda^A,\lambda'^B \in \Lambda_{\ket{0}} \subset \Lambda^B$ and $\lambda'^C \in \Lambda_{\ket{0}} \subset \Lambda^C$. On the outputs GHZ test is performed. If, instead of $\ket{0}$, $\ket{+}$ is fed as input, GHZ test yields maximal Mermin violation of $4$ - which would be impossible if $\Lambda_{\ket{0}}$ and $\Lambda_{\ket{+}}$ have nonzero overlap. }\label{fig.pdf}
\end{figure}

We now analyze the action of $\mathbb{M}$ at the ontological level. Whenever the $1$st port is fed with system $A$ prepared in the state $\ket{0}$, the machine receives a composite ontic state $\lambda^{ABC}= (\lambda^A,\lambda^B,\lambda^C)$ where $\lambda^A \in \Lambda_{\ket{0}}\subset \Lambda^A,~\lambda^B\in \Lambda_{\ket{r}}\subset\Lambda^B$ and $\lambda^C \in \Lambda_{\ket{r}} \subset \Lambda^C$. Receiving the inputs from its three ports, the machine yields the outcome $\ket{0}_A\ket{0}_B\ket{0}_C$  {\it i.e.} at the output it yields a joint ontic state $\mathbb{M}[\lambda^{A},\lambda^B,\lambda^C] = (\lambda'^A,\lambda'^B,\lambda'^C)$ where $\lambda'^A \in \Lambda_{\ket{0}}$, $\lambda'^B \in \Lambda_{\ket{0}}$ and $\lambda'^C \in \Lambda_{\ket{0}}$ (see Figure \ref{fig.pdf}). Therefore, the output $\mathbb{M}[\lambda^{A},\lambda^B,\lambda^C] = (\lambda'^A,\lambda'^B,\lambda'^C)$ belongs in $ \Lambda_{\ket{0}} \times \Lambda_{\ket{0}} \times \Lambda_{\ket{0}} \subset \Lambda^A \times \Lambda^B \times \Lambda^C $. Hence, the output $\mathbb{M}[\lambda^{A},\lambda^B,\lambda^C]$ is local - the Mermin value $\mathcal{M}$ of the output $\mathbb{M}[\lambda^{A},\lambda^B,\lambda^C]$ is bounded by $2$,  where $\mathcal{M} := \langle a_0 b_0 c_1 \rangle + \langle a_0 b_1 c_0 \rangle + \langle a_1 b_0 c_0 \rangle - \langle a_1 b_1 c_1 \rangle$ and $\{a_0,a_1\}$, $\{b_0, b_1\}$, $\{c_0,c_1\}$ are dichotomic local measurements with outcomes $\{+1,-1\}$ on $A$, $B$, $C$ respectively. \cite{Mermin90,Scarani}. So we note machine's action : [$\Gamma$] whenever $\lambda^A$ from $\Lambda_{\ket{0}}$ is fed as input in the $1$st port of $\mathbb{M}$, the output $\mathbb{M}[\lambda^{A},\lambda^B,\lambda^C]$ satisfies  $\mathcal{M}_{\mathbb{M}[\lambda^{A},\lambda^B,\lambda^C]} \leq 2$ 

Let us now feed the $1$st port of $\mathbb{M}$ with the system A prepared in state $\ket{+}=(\ket{0}+\ket{1})/\sqrt{2}$. At the ontological level, the machine receives some joint ontic states $\lambda^{ABC} = (\lambda^A,\lambda^B,\lambda^C)$ where $\lambda^A \in \Lambda_{\ket{+}}$ $\lambda^B \in \Lambda_{\ket{r}}$ $\lambda^C \in \Lambda_{\ket{r}}$ sampled with probability distribution $\mu(\lambda^A,\lambda^B,\lambda^C|+rr)$. Assume that the ontic supports $\Lambda_{\ket{0}}$ and $\Lambda_{\ket{+}}$ have nonzero overlap (see Figure \ref{fig.pdf}).  Therefore, whenever machine receives $\lambda^A$ from the overlap $\Lambda_{\ket{0}} \cap \Lambda_{\ket{+}}$, machine's action on such $\lambda^{ABC} \equiv (\lambda^A,\lambda^B,\lambda^C)$ is already known (from [$\Gamma$]) : for $\lambda^A \in \Lambda_{\ket{+}} \cap \Lambda_{\ket{0}}$, $\mathcal{M}_{\mathbb{M}[\lambda^A,\lambda^B,\lambda^C]} \leq 2 $.

But due to linearity of the machine's action one will obtain the output $\ket{GHZ}=(\ket{0}\ket{0}\ket{0}+\ket{1}\ket{1}\ket{1})/\sqrt{2}$ whenever the machine $\mathbb{M}$ is fed with state $\ket{+}$. The tripartite state $\ket{GHZ}$ exhibits maximum violation $4$ of Mermin inequality \cite{Mermin90,Scarani} for suitable measurement choice on its local parts. Now quantum reproducibility condition for
the observed nonlocality demands,

\footnotesize
\begin{align}
&\int_{\Lambda_{\ket{+}}} d\lambda^A  \int_{\Lambda_{\ket{r}}} \int_{\Lambda_{\ket{r}}} d\lambda^B d\lambda^C \mu(\lambda^{A},\lambda^{B},\lambda^{C}|+rr)\mathcal{M}_{\mathbb{M}[\lambda^{A},\lambda^{B},\lambda^{C}]}\nonumber\\
&= \bra{GHZ}\mathcal{M}\ket{GHZ}= 4 \label{eq1}
\end{align}
\normalsize

 The domain of integration for $\lambda^A$ can be divided in two disjoint parts : $\Lambda_{\ket{+}} = (\Lambda_{\ket{+}} \cap \Lambda_{\ket{0}}) \cup (\Lambda_{\ket{+}} \setminus (\Lambda_{\ket{+}} \cap \Lambda_{\ket{0}})) $. As argued above, $\mathcal{M}_{\mathbb{M}[\lambda^A,\lambda^B,\lambda^C]} \leq 2$ whenever $\lambda^A \in \Lambda_{\ket{+}} \cap \Lambda_{\ket{0}}$. Thus we have, 

\footnotesize
\begin{align}
&\int_{\Lambda_{\ket{+}} \cap \Lambda_{\ket{0}}} d\lambda^A  \int_{\Lambda_{\ket{r}}} \int_{\Lambda_{\ket{r}}}d\lambda^B d\lambda^C \mu(\lambda^{A},\lambda^{B},\lambda^{C}|+rr)\mathcal{M}_{\mathbb{M}[\lambda^{A},\lambda^{B},\lambda^{C}]}\nonumber\\
&=\int_{\Lambda_{\ket{+}}\cap\Lambda_{\ket{0}}} d\lambda^A \int d\lambda^B d\lambda^C \mu(\lambda^A,\lambda^B,\lambda^C|+rr)\times 2 \nonumber\\ 
&=2\int_{\Lambda_{\ket{+}} \cap \Lambda_{\ket{0}}} d\lambda^A\mu(\lambda^A|+)\nonumber \\
&=2\left[\int_{\Lambda_{\ket{0}}} d\lambda^A \mu(\lambda^A|+) - \int_{\Lambda_{\ket{0}}\setminus (\Lambda_{\ket{+}} \cap \Lambda_{\ket{0}})} d\lambda^A(\mu(\lambda^A|+) \right] \nonumber\\
&=2\left[\int_{\Lambda_{\ket{0}}} d\lambda^A \mu(\lambda^A|+) - 0 \right]\nonumber\\
&= 2 \Omega(0,+)|\langle 0 |{+}\rangle|^2 =\Omega(0,+). \label{eq2} 
\end{align}
\normalsize
We have considered that local ontic states yield the maximum possible value $2$ for the Mermin expression $\mathcal{M}$ and used the fact that $\int d\lambda^B d\lambda^C \mu(\lambda^A,\lambda^B,\lambda^C|+rr)=\mu(\lambda^A|+)$ -  here it should be clear to the reader  that this step is obtained simply as any marginal probability distribution (here $\mu(\lambda^A|+)$) is obtained from the joint distribution (here $\mu(\lambda^A,\lambda^B,\lambda^C|+rr)$) by summing/integrating over other variables. For this to hold, the joint distribution $\mu(\lambda^A,\lambda^B,\lambda^C|+rr)$ does not need to be product of its marginals.

Whenever, $\lambda^{A}\in\Lambda_{\ket{+}}\setminus (\Lambda_{\ket{+}}\cap \Lambda_{\ket{0}})$, the output $\mathbb{M}[\lambda^A,\lambda^B,\lambda^C]$ may lie in $\Lambda^{NL}$ and contributes to the observed quantum nonlocality. Assuming that all such $\mathbb{M}[\lambda^A,\lambda^B,\lambda^C]$ yield the maximum possible Mermin value $\mathcal{M}$ ({\it i.e.} $4$) we obtain

\footnotesize
\begin{align}
&\int_{\Lambda_{\ket{+}}\setminus(\Lambda_{\ket{0}}\cap\Lambda_{\ket{+}})} d\lambda^A \int d\lambda^B d\lambda^C \mu(\lambda^A,\lambda^B,\lambda^C|+rr) \mathcal{M}_{\mathbb{M}[\lambda^A,\lambda^B,\lambda^C]}\nonumber\\
&~=\int_{\Lambda_{\ket{+}}\setminus(\Lambda_{\ket{0}}\cap\Lambda_{\ket{+}})} d\lambda^A \int d\lambda^B d\lambda^C \mu(\lambda^A,\lambda^B,\lambda^C|\ket{+rr})\times 4\nonumber\\
&=4\left[\int_{\Lambda_{\ket{+}}} d\lambda^A\mu(\lambda^A|+)-\int_{\Lambda_{\ket{0}}} d\lambda^A\mu(\lambda^A|+) \right]\nonumber\\
&=4\left[1-|\langle 0|+ \rangle|^2\Omega(0,+) \right] \nonumber\\
&= 4\left[1-\frac{1}{2}\Omega(0,+) \right]\label{eq4}
\end{align}
\normalsize
Eqs. (\ref{eq1}), (\ref{eq2}), (\ref{eq4}) together imply
\footnotesize
\begin{align}
\Omega(0,+) + 4\left[1-\frac{1}{2}\Omega(0,+) \right] = 4 \label{eq5}
\end{align}
\normalsize

Therefore, $\Omega(0,+) = 0 $ {\it i.e.} there is no overlap between $\Lambda_{\ket{0}}$ and $\Lambda_{\ket{+}}$. Thus the quantum indistinguishability of $\ket{0}$ and $\ket{+}$ cannot be explained by overlap in their respective Liouville distributions. Models where indistinguishabiltiy is attributed, even partially, to overlaps between Liouville distributions cannot succeed in reproducing the maximal violation of Mermin inequality in the above experiment. In fact, it is easy to see that the conclusion $\Omega(\psi,\phi) = 0$ holds in general for any pair of $\ket{\psi}$ and $\ket{\phi}$ with $\langle \psi | \phi \rangle = \frac{1}{\sqrt{2}}$. Thus the result also rules out maximally $\psi$-epsitemic interpretations even for dimension two (compare with \cite{Leifer13}).

It is worth emphasizing here that the proof does not require the preparation independence assumption of PBR. The proof runs equally well even if we assume $\lambda^A$, $\lambda^B$ and $\lambda^C$ are perfectly correlated. Rather, what is crucial to the proof is the following : for a given $\lambda^A \in \Lambda^A_{\ket{0}} \cap \Lambda_{\ket{+}}$, the joint ontic state $\lambda^{ABC} = (\lambda^A,\lambda^B,\lambda^C)$ that goes as input when that $\lambda^A$ is fed as a $\ket{0}$ preparation also goes as input when that particular $\lambda^A$ is fed as a $\ket{+}$ preparation. This is natural since $\lambda^A$ belongs in the overlap of $\Lambda_{\ket{0}}$ and $\Lambda_{\ket{+}}$, and therefore realisation of $\lambda^A$ does not contain the information whether it is a $\ket{0}$ preparation or $\ket{+}$ preparation. If it was otherwise, complete characterisation of such $\lambda^A$ would require to be supplemented with the additional information of the quantum state which it gets realised as ( {\it i.e.} realisation of $ \lambda_0^A : = (\lambda^A,\ket{0})$ and of $\lambda_+^A : = (\lambda^A,\ket{+})$ for the same $\lambda^A \in \Lambda_{\ket{0}} \cap \Lambda_{\ket{+}}$ would have to be considered two distinct ontological phenomena). Such 'quantum state-supplemented' (also called $\psi$-supplemented) descriptions of state of reality can only be $\psi$-ontic \cite{Harrigan09}, which would ultimately imply zero overlap between ontic supports of distinct quantum states.

{\it Discussions:--} We have given a proof that in any ontological model, certain pairs of non-orthogonal quantum states must have zero overlap in their epistemic distributions. The fact that maximal violation of Mermin inequality is attainable in quantum theory cannot be explained otherwise. This leads to the more delicate issue of how we should interpret indistinguishability of non-orthogonal quantum states. Two possible explanations still survive. If the model is deterministic, distinct quantum states whose ontic supports do not overlap could in principle be distinguished in a single shot unless there is some restriction on possibility of joint measurement of different observables. Thus for determinsitic ontological models, indistinguishability must be attributed to the impossibility of sharp joint measurement. The other possible explanation is that the model has to be indeterministic {\it i.e.} the ontic state $\lambda$ that contain complete information of the system yields only probability - the standard quantum theory is itself one such example. Though there are a number of results that indicate towards the same direction as concluded in this paper \cite{Pusey12,Barrett14,Hardy13}, the strength of the present theorem lies in several aspects. Firstly, its conclusion does not rely on assumptions like 'preparation independence' used by PBR or 'Ontic indifference' by Hardy \cite{Hardy13}. Secondly, it not only puts restriction on the amount of ontic overap between two nonorthogonal states, rather it dictates zero overlap between them. It is worth mentioning here that Leifer {\it{et al}} showed maximally $\psi$-epsitemic model $\Rightarrow$ Kochen-Specker noncontextuality \cite{Leifer13}. Since Kochen-Specker theorem applies to Hilbert space dimension strictly greater than two \cite{Kochen67}, their result cannot discard maximally $\psi$-epistemic interpretations for dimension two.  However, due to our theorem the maximally $\psi$-epsitemic interpretation gets ruled out also for dimension two. While we have reached the conclusion using maximal violation of Mermin inequality, it would be interesting to see what can be concluded regarding the possible interpretations of quantum indistinguishability using Svetlichny inequality \cite{Svetlichny87}. It would also be interesting to construct a test for experimental certification of the conclusion derived in this paper. 

\begin{acknowledgments}
ADB acknowledges fruitful discussions with Guruprasad Kar and Manik Banik.
\end{acknowledgments}

\end{document}